\documentclass[twocolumn]{revtex4}
\usepackage{graphicx}

\begin{document}

\title{Unconventional Superconductivity in Heavy Fermion Systems}

\author{Y. \textsc{Kitaoka}, S. \textsc{Kawasaki}, 
T. \textsc{Mito}\thanks{Present address: Department of Physics, Faculty of Science, Kobe University, Nada, Kobe 657-8501, Japan}, Y. \textsc{Kawasaki}\thanks{Present address: Department of Physics, Faculty of Engineering, Tokushima University, Tokushima 770-8506, Japan}}

\affiliation{Department of Materials Engineering Science, Graduate School of Engineering Science, Osaka University, Toyonaka, Osaka 560-8531, Japan}%

\date{\today}

\begin{abstract}
We review the studies on the emergent phases of superconductvity and magnetism in the $f$-electron derived heavy-fermion (HF) systems by means of the nuclear-quadrupole-resonance (NQR) under pressure. These studies have unraveled a rich variety of the phenomena in the ground state of HF systems. In this article, we highlight the novel phase diagrams of magnetism and unconventional superconductivity (SC) in CeCu$_2$Si$_2$, HF antiferromagnets CeRhIn$_5$, and CeIn$_3$. A new light is shed on the difference and common features on the interplay between magnetism and SC on the magnetic criticality.
\end{abstract}

\keywords{Kondo lattice, heavy-fermion superconductivity, pressure-induced phase transition, quantum critical point, NQR}

\maketitle
\section{Introduction}
Many of lanthanide (Ce, Yb, etc) and actinide (U, Pu, etc) compounds behave as if they consist of an assembly of local moments at temperatures much higher than the characteristic energy of hybridization or intersite exchange interaction. This quasi-independent behavior continues sometimes down to low temperatures.
In the case where the hybridization energy is larger than the crystal electric field (CEF) level splittings, the splittings are smeared out.  
Then the large degeneracy $n$ associated with the $J$ multiplet plays an important role in determining the characteristic energy scale which is typically 100 K.  
For example we have $n=6$ for Ce$^{3+}$ and $n=8$ for Yb$^{3+}$.
In many lanthanide compounds, the valence fluctuating state has been probed by means of X-ray photoemission spectroscopy, measurements of lattice constant and M\"{o}ssbauer isomer shift.
In the valence fluctuation regime, the strong hybridization effect leads to both charge and spin fluctuations. As a consequence the  number $n_{f}$ of $f$ electrons is less than one.

On the contrary, if the hybridization between the $f$- and conduction electrons is relatively weak, the charge
fluctuation is suppressed since the Coulomb repulsion
 between $f$ electrons is strong.  In the latter case the hybridization generates the exchange interaction $J_{cf}$ between $f$- and conduction electron spins.  This case is referred to as the Kondo regime \cite{Kondo,Fulde,Hewson}. The characteristic energy scale, $T_K$, in the Kondo regime is typically 10 K. In most cases, the CEF splitting  becomes larger than  $T_{K}$.

At high temperatures, an assembly of local moments behaves as independent Kondo scattering centers.  For example, the Curie-Weiss behavior of the susceptibility appears and the $\log{T}$-anomalies of the resistivity take places. Therefore, various aspects of the Kondo effect including the dynamical characteristics can be investigated by means of nuclear-magnetic resonance(NMR) and neutron scattering experiments of sufficient accuracy. 
The difference in the characteristic energy scales between the valence fluctuation and the Kondo regimes manifests itself in the $T$ dependence of the magnetic relaxation rate. If the dynamical susceptibility is isotropic and can be approximated by a Lorentzian with the magnetic relaxation rate $\Gamma$, nuclear spin lattice relaxation rate $1/T_{1}$ is expressed as
\begin{equation}
\frac{1}{T_{1}}=2\gamma_{n}^{2} k_{B}T|A_{hf}|^{2} \frac{\chi(T)}{\Gamma},
\label{eq:LT1}
\end{equation}
where $A_{hf}$ is the average hyperfine field.
For quasielastic neutron scattering, the magnetic cross
section is derived  as
\begin{equation}
\frac{d^{2}\sigma}{d\Omega d\omega} \sim A^{2}\frac{k_{1}}{k_{0}}
 | F(\mbox{\boldmath$q$})| ^{2}\chi(T)
\frac{\omega}{1-\exp(-\omega/k_{B}T)}\frac{\Gamma}{\Gamma^{2}+\omega^{2}}.
\label{eq:Nquasi-elastic}
\end{equation}
Thus $\Gamma$ appears as the half width in the quasielastic neutron scattering spectrum.
In the limit of small $\omega$, the imaginary part of $\chi(\omega)$ obeys the Korringa-Shiba relation \cite{Shiba75}.
Accordingly, with the $T_1T$=constant law at low temperatures we can estimate $\Gamma$ by the relation
\begin{equation}
\Gamma =\frac{2
\gamma_{n}^{2}T_{1}T\chi(0)|A_{hf}|^{2}}{\mu_{B}^{2}}.
\label{eq:EXgamma}
\end{equation}

The NMR relaxation rate and the half-width of the quasi-elastic magnetic neutron scattering spectrum can be understood in a consistent way.
The width of the quasi-elastic spectrum $\Gamma$ is a
measure of the strength of the hybridization or the exchange interaction between the $f$- and the conduction electrons. In rare earth compounds with a magnetically stable $4f$ configuration, one expects a Korringa behavior for the quasi-elastic linewidth, namely $\Gamma=\alpha k_{B}T$ where $\alpha$ is
typically $10^{-3}$, while the valence fluctuation compounds show an almost $T$ independent $\Gamma$, reaching typically to
 $20 \sim 30$ meV.
 
In contrast to these, the relaxation rate in the Kondo regime 
exhibits a characteristic $T$ dependence and probes the presence of a very low energy scale of $10 \sim 30$ K. Then the presence of such a low energy scale is related to the huge
linear term of the specific heat which amounts to about 1 J/moleK$^{2}$.  Namely the system can be described by renormalized heavy quasi-particles at low temperatures.  For antiferromagnets with heavy 
quasi-particles, quasi-elastic Lorentzian intensities still survive even below a magnetic transition temperature $T_{M}$ and then, show a distinct deviation from a $T$-linear dependence. In most cases, $\Gamma(T) $ seems to follow roughly a square-root dependence of $\Gamma=A \sqrt{T}$ \cite{Maekawa}.

We would say that the spin dynamics probed by NMR and neutron scattering techniques can be semiquantitatively interpreted within such impurity models as the degenerate Anderson \cite{Anderson61} and the Coqblin-Schrieffer \cite{Coqblin} models as long as high temperature behaviors are concerned.
In this article, we shall see how the strong interaction effects among electrons in {\it Kondo lattice compounds} lead to richer and more fascinating behaviors in {\it heavy-fermion} (HF) state at low temperatures via the microscopic probe by NMR and nuclear-quadrupole resonance(NQR).
\section{Experimental Details}
For understanding specific properties of HF systems,
it is important to investigate and describe low energy excitations. In this context, both NMR/NQR and neutron scattering experiments play central roles in obtaining information about the dynamical response of HF systems.
Whereas the NMR/NQR probes the local environment of one particular nucleus and, therefore, a wave-vector average of the dynamical response function with a small energy transfer comparable to the nuclear Zeeman energy ($10^{-5}\sim 10^{-3}$ meV), neutron scattering experiments can scan wider energies and wave vectors corresponding to the whole Brillouin zone. In real situations the energy transfer is usually limited below about 100 meV with use of thermal neutrons. Thus, both experiments are complementary. The advantage of NMR/NQR
 is that it can extract the lowest energy excitation and detect a
magnetic instability with a tiny moment if it appears. Furthermore in the superconducting state, NMR/NQR can provide a detailed structure of response function reliably.
For more detailed descriptions on NMR/NQR, we refer to the textbooks \cite{NMRREF}.
\section{Formation of Heavy-Fermion Metals with Magnetic Correlations}
Magnetic and thermal properties of heavy fermions are basically determined by strong local correlation.  A signature of the local nature is that the susceptibility and specific heat are roughly proportional to concentration of Ce in systems such as Ce$_{x}$La$_{1-x}$Cu$_{6}$ or Ce$_{x}$La$_{1-x}$B$_{6}$.\cite{Onuki1987} The RKKY interaction is operative in bringing about magnetic ordering.  The energy scale of it is given by $J_{RKKY}\sim J^2\rho_c$ where $J$ is the exchange interaction between $f$ and conduction electrons and $\rho_c$ is the density of conduction band states per site. The RKKY interaction competes with the Kondo effect which points to the non-magnetic singlet ground state. There exists a rather detailed balance between these two tendencies. It is expected that $T_{M}$, if it exists, is lowered due to the Kondo effect.  In the opposite case of $T_K\gg J^2\rho_c$, no magnetic order should occur.

Even in the singlet ground state, intersite interaction should influence residual interactions among renormalized heavy quasi-particles. It is this residual interaction that leads to a rich variety in the ground states such as HF superconductivity, HF band magnetism, etc. Short-range magnetic correlation might survive even if the dominating Kondo effect prevents development of long-range order. There, electrical resistivity reveals growing importance of coherence upon increasing concentration of rare-earth ions at low temperatures. 

Direct evidence for heavy quasi-particles comes from the de Haas-van Alphen (dHvA) effect that is the oscillation of the magnetic susceptibility coming from Landau quantization of the electron orbits \cite{Shoenberg}. Geometry of the Fermi surface can be obtained from the period of oscillation in the differential magnetic susceptibility as a function of $1/H$ with $H$ the magnetic field. The effective mass can be deduced from $T$ dependence of the oscillation amplitude.
To observe these oscillations, the mean free path $l$ of a heavy quasiparticle must be larger than the cyclotron radius: $l\gg v_F/\omega_H$ where $v_F$ is the Fermi velocity, and $\omega_H=eH/m^*c$ is the effective cyclotron frequency of a heavy quasi-particle with effective mass $m^*$.
Furthermore thermal smearing at the Fermi surface must be sufficiently small: $T\ll\omega_H$.  With large effective masses and short mean free paths, these conditions require that  experiment should be performed at low temperature and in high field. Measurements of dHvA oscillations have been made on UPt$_3$ \cite{de-HaasUPt3,Julian}, CeRu$_2$Si$_2$ \cite{de-HaasCeRu2Si2,Julian} and CeCu$_6$ \cite{de-HaasCeCu6}. 
In all these cases,  heavy effective masses have been observed. 

For UPt$_3$,  the Fermi surface obtained by the dHvA effect is in good agreement with that obtained from energy-band theory \cite{band-theory}. The effective mass on some sheets of the Fermi surface, however, is about 10 to 30 times larger than that deduced from band-structure calculations. This fact shows that there are significant many body effects that are not described by the standard band theory.

Here, we shall review paramagnetic HF states that have metallic phase. In the following we focus on the growth of intersite coherence, or the itinerant character of $f$ electrons, mainly in dynamic magnetic properties. As specific examples to see the subtle role of the intersite correlation in
the HF state, we take typical systems CeCu$_{6}$ and
CeRu$_{2}$Si$_{2}$ where the Fermi-liquid remains stable at low
temperatures even with significant magnetic correlations.

\begin{figure}[h]
\centering
\includegraphics[width=7cm]{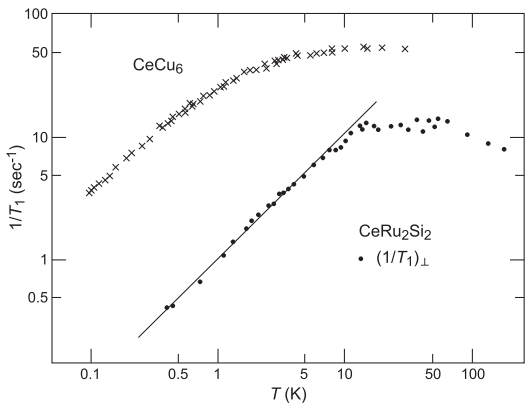}
\caption[]{\footnotesize Temperature dependences of $(1/T_1)$ of $^{63}$Cu in
CeCu$_{6}$($\times$) and $^{29}$Si in
CeRu$_{2}$Si$_{2}$($\bullet$).\cite{Kitaoka1987} Solid line denotes a $T_1T$=constant law.}
\label{fig1_T1_CeCu6}
\end{figure}

Figure \ref{fig1_T1_CeCu6} shows the $T$ dependence of $1/T_{1}$ for CeCu$_{6}$ and CeRu$_{2}$Si$_{2}$.\cite{Kitaoka1987}
In the latter case $1/T_{1\perp}$ of $^{29}$Si is measured under the condition that the c-axis is aligned along the magnetic field. For CeCu$_{6}$, $1/T_{1}$ of $^{63}$Cu was measured in zero field by Cu NQR. As seen in the figure, $1/T_{1}$ is almost independent of temperature above 6 K for CeCu$_{6}$ and 12 K for CeRu$_{2}$Si$_{2}$. Then $1/T_{1}$ begins to decrease gradually above 50 K for CeRu$_{2}$Si$_{2}$. This relaxation behavior shares a common feature with other HF systems at high temperatures. Spin correlations between different sites can be ignored at high temperatures, and $1/T_{1}$ is described by the local spin susceptibility $\chi(T)$ and the magnetic relaxation rate $\Gamma$.

\begin{figure}[htbp]
\centering
\includegraphics[width=7cm]{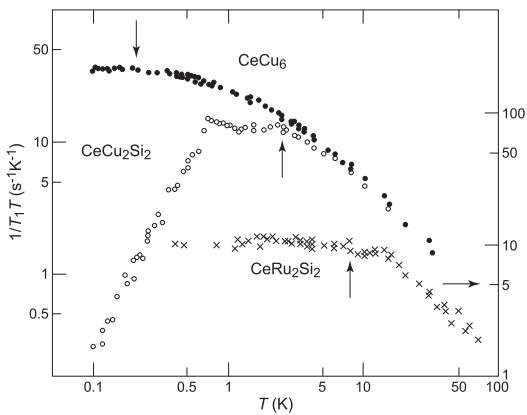}
\caption[]{\footnotesize Temperature dependences of $1/(T_{1}T)$  of $^{63}$Cu in CeCu$_{6}$($\bullet$) and CeCu$_{2}$Si$_{2}$($\circ$) and $^{29}$Si in CeRu$_{2}$Si$_{2}$($\times$).\cite{Kitaoka1987} Arrows denote $T_{FL}$ below which $1/T_1$ follows a behavior $T_1T$=constant. The $1/T_1T$ in the HF superconductor CeCu$_{2}$Si$_{2}$($\circ$) follows a behavior $1/T_1\propto T^3$ below $T_c$, revealing the presence of the line node in energy-gap structure.}
\label{fig2_T1T_CeCu6}
\end{figure}

The temperature below which $1/T_{1}$ begins to decrease is
close to the Kondo temperature $T_{K}$ which is extracted from analysis of the resistivity and the magnetic specific heat.
With further decrease of temperature, $1/T_{1}$ follows
a behavior $T_{1}T$=constant, which is called the Korringa
law. To see this clearly, Fig.\ref{fig2_T1T_CeCu6} shows $(T_{1}T)^{-1}$ vs $T$.
It turns out that the Korringa law is valid below $T_{FL}$=0.2, 3 and 8 K for CeCu$_{6}$, CeCu$_2$Si$_2$ and CeRu$_{2}$Si$_{2}$, respectively. The characteristic temperature $T_{FL}$ in each case is defined as the effective Fermi liquid temperature. The Fermi liquid ground state leads to the Korringa law at temperatures lower than $T_{FL}$. This is independent of whether the system is homogeneous or a local Fermi liquid. In the temperature range where the local Fermi liquid description is valid, the Korringa relation should hold.

In contrast to the expectation based on the single-ion model of spin fluctuations, the ratio of $T_{FL}/T_{K}$ quite depends on systems; $T_{FL}/T_{K}$=0.2/6=0.03, 3/10=0.3 and  8/12=0.67 for CeCu$_{6}$, CeCu$_2$Si$_2$ and CeRu$_{2}$Si$_{2}$, respectively.
Furthermore $T_{FL}$ decreases to 0.08 K in Ce$_{0.75}$La$_{0.25}$Cu$_{6}$.\cite{Onuki1987} These results show that the energy scale $T_{K}$ is not sufficient to describe the HF systems.  It should also be specified by the effective Fermi temperature, $T_{FL}$ below which HF bands are formed.
In contrast to the universal behavior of the Kondo impurity, the dispersion relations of heavy quasi-particles depend on materials.\\
\section{Unconventional Superconductivity in Heavy-Fermion Systems}
\subsection{Overview}
Superconductivity, which was one of the best understood many-body problems in physics, became again a challenging problem when a new kind of superconductivity was discovered in CeCu$_{2}$Si$_{2}$ by Steglich et al.\cite{Steglich1979} The system is one of HF materials close to magnetic instability. In the subsequent decade, intensive investigations of a class of uranium compounds established a new field of HF superconductivity by successive discoveries of superconductivity in UBe$_{13}$,
UPt$_{3}$, URu$_{2}$Si$_{2}$,
UPd$_{2}$Al$_{3}$
and UNi$_{2}$Al$_{3}$.\cite{SCES94,Grewe1991}  
The most important characteristics for a series of uranium  HF superconductors  are that superconductivity coexists with the antiferromagnetism except for
UBe$_{13}$, and that the  specific heat coefficient $\gamma$ lies in a broad range from 700 mJ/mole K$^{2}$ (UBe$_{13}$) to 60 mJ/mole K$^{2}$ (URu$_{2}$Si$_{2}$). The $f$-shell electrons, which are strongly correlated by Coulomb repulsive interaction, determine the properties of heavy quasi-particles at the Fermi level. This gives rise to a large $\gamma$ value as well as an enhanced spin susceptibility. Hence, the Fermi energy is also quite small: $T_{FL}=10\sim 100$K and, as a result, the transition temperature is also small, ranging from $T_{c}=0.5$K to 2K.  
The magnetic ordering temperature $T_{N}=5\sim 20$K is by one order of magnitude higher than $T_{c}$. A jump $(C_{s}-C_{n})/C_{n}$ of the specific heat normalized by the value $C_{n}$ just above $T_{c}$ is of $O(1)$ in all compounds.
This result demonstrates that the superconductivity is produced mainly by the heavy quasi-particles. Due to the strong Coulomb repulsion among $f$ electrons, it seems hard for the heavy quasi-particles to form ordinary $s$-wave Cooper pairs with large amplitude at zero separation of the pair. In order to avoid Coulomb repulsion, the system would favor an anisotropic pairing channel like spin triplet $p$-wave or spin singlet $d$-wave. 

As listed in Table 1, the HF superconductors discovered to date may be classified into two groups depending on their different magnetic behaviors and extent of quasi-particle renormalization. The first group comprises CeCu$_{2}$Si$_{2}$, CeIrIn$_5$, CeCoIn$_5$, UBe$_{13}$, and UPt$_{3}$, which exhibit no magnetic order or quasi-static order. The quasi-particle masses as derived from the specific-heat coefficient $\gamma=C(T)/T$ are as large as $\gamma \geq 400$ mJ/mole K$^{2}$.
The highest $T_{c}$ is 2.3 K in CeCoIn$_5$ \cite{Petrovic1,Petrovic2}.
\begin{widetext}

\begin{center}
\begin{table}[htbp]
\begin{tabular}{l|c|c|c|c|c|c|c}
\hline
\hline
 & $T_c$(K) & crystal structure & nucleus & 1/$T_1$ & $K^*$ & parity & symmetry \\ 
\hline
CeCu$_2$Si$_2$ \cite{Steglich1979,Bellarbi,Bellarbi2,kawasaki01,Ykawasaki2} & $\sim$ 0.7 K & tetragonal(ThCr$_2$Si$_2$) & Cu, Si \cite{Ueda,Ishida1} & $T^3$ & decrease & even & $d$ \\
CeCoIn$_5$ \cite{Petrovic1,Petrovic2} & $\sim$ 2.3 K & tetragonal(HoCoGa$_5$) & Co, In \cite{Kohori} & $T^3$ & decrease & even & $d$ \\
CeIrIn$_5$ \cite{Petrovic1,Petrovic2} & $\sim$ 0.4 K & tetragonal(HoCoGa$_5$)  & In \cite{Zheng} & $T^3$ & - & - & - \\
UBe$_{13}$ \cite{SCES94,Grewe1991}  & $\sim$ 0.9 K & cubic(NaZn$_{13}$) & Be \cite{Tien} & $T^3$ & - & - & - \\
UPt$_3$  \cite{SCES94,Grewe1991} & $\sim$ 0.55 K & hexagonal & Pt  \cite{Kohori1,Kohori1_2,Tou1,Tou1_2} & $T^3$ & unchange & odd & $p$ or $f$ \\
\hline
URu$_2$Si$_2$  \cite{SCES94,Grewe1991} & $\sim$ 1.2 K & tetragonal(ThCr$_2$Si$_2$) & Ru, Si \cite{Kohori2,Matsuda} & $T^3$ & unchange & odd &  \\
UNi$_2$Al$_3$  \cite{SCES94,Grewe1991} & $\sim$ 1 K & hexagonal & Al \cite{Ishida2} & $T^3$ & unchange & odd & $p$ or $f$  \\
UPd$_2$Al$_3$  \cite{SCES94,Grewe1991} & $\sim$ 2 K & hexagonal & Pd, Al \cite{Kyogaku,Tou2} & $T^3$ & decrease & even & $d$  \\
\hline
CeCu$_2$Ge$_2$ \cite{Jaccard92} & $\sim$ 0.6 K ($P\sim$7.6 GPa) & tetragonal(ThCr$_2$Si$_2$) & - & - & - & - & - \\
CeIn$_3$ \cite{Mathur98,Grosche01,Walker97,Muramatsu01,Knebel02} & $\sim$ 0.2 K ($P\sim$2.5 GPa) & cubic(AuCu$_3$) & In \cite{Shinji04}& $T^3$ & - & - & - \\
CePd$_2$Si$_2$  \cite{Mathur98,Grosche96,Grosche01} & $\sim$ 0.4 K ($P\sim$2.5 GPa) & tetragonal(ThCr$_2$Si$_2$) & - & - & - & - & - \\
CeRh$_2$Si$_2$ \cite{Movshovic96,Araki} & $\sim$ 0.2 K ($P\sim$1.0 GPa) & tetragonal(ThCr$_2$Si$_2$) & - & - & - & - & - \\
CeRhIn$_5$ \cite{Hegger00,Muramatsu} & $\sim$ 2.1 K ($P\sim$1.6 GPa) & tetragonal(HoCoGa$_5$)  & In  \cite{Mito01,Kohori4} & $T^3$ & - & - & - \\
\hline
\hline
High-$T_c$ cuprates & $\sim$ 140 K (max) & perovskite & Cu, O & $T^3$ & decrease & even & $d$  \\
\hline
Sr$_2$RuO$_4$ \cite{Ishida3,Ishida3_2} & $\sim$ 1.5 K & perovskite & Ru, O & $T^3$ & unchange & odd & $p$ \\ 
\hline
\hline
\end{tabular}

\caption[]{Superconducting characteristics in most heavy-fermion systems along with high-$T_c$ cupper oxides and Sr$_2$RuO$_4$. Note that the nuclear relaxation rate $1/T_1$ reveals no coherence peak just below $T_c$, followed by the $T^3$ dependence without an exception. $K^*$ denotes the spin component of Knight shift below $T_c$. In this context,  all unconventional superconductors discovered to date possess the line-node gap on the Fermi surface regardless of either spin-singlet $d$ wave or spin-triplet $p$-wave.}
\label{para}
\end{table}
\end{center}
\end{widetext}
The second group includes URu$_{2}$Si$_{2}$, UNi$_{2}$Al$_{3}$ and UPd$_{2}$Al$_{3}$, which show antiferromagnetic order.
The effective masses are not so large with $\gamma \sim$
100 mJ/mole K $^{2}$. Most remarkably, UNi$_{2}$Al$_{3}$ and UPd$_{2}$Al$_{3}$ possess
sizable magnetic moments of 0.24  and 0.85 $\mu_{B}$/(U atom),
respectively. In both groups, heavy quasi-particles are responsible
for the superconductivity.  This follows from the large jump of the
specific heat $\Delta C$  at $T_{c}$ with $\Delta C/\gamma T_{c}\sim 1$, which is close to the value 1.43 for conventional superconductors, and from the large slope of the upper critical field $(dH_{c2}/dT)_{T=T_{c}}$.
\subsection{NMR as a probe of superconducting states}
Nuclear spin-lattice relaxation rate $1/T_1$ is expressed as,
\[
\frac{1}{T_1T}=\frac{\pi A_{hf}^2\alpha (T)}{\hbar}\int dE\left\{N_s^2(E)+M_s^2(E)\right\}\left\{-\frac{\partial f(E)}{\partial E}\right\}.\]
Here $\alpha(T)$ expresses an enhancement factor due to electron correlation effect. The $T$ dependence of $\alpha(T)$ is neglected for the present. $A_{hf}$ is the hyperfine-coupling constant. $E$ and $f(E)$ are the energy of the quasi-particles and the Fermi function, respectively. The density of states of the quasi-particles $N_s(E)$ and the anomalous density of states $M_s(E)$ associated with the coherence effect are expressed as follows:
\begin{equation}
N_s(E)=\frac{N_0}{4\pi}\int_\Omega\frac{E}{\sqrt{E^2-\Delta^2(\theta,\phi )}}d\Omega, 
\label{DOS1}
\end{equation}
\begin{equation}
M_s(E)=\frac{N_0}{4\pi}\int_\Omega\frac{\Delta(\theta,\phi )}{\sqrt{E^2-\Delta^2(\theta,\phi)}}d\Omega
\label{DOS2}
\end{equation}
For the $s$-wave with isotropic gap $\Delta_0$, the respective eqs. (\ref{DOS1}) and (\ref{DOS2}) are simplified to 
\[N_s(E)=N_0\frac{E}{\sqrt{E^2-\Delta_0^2}}, \hspace{5mm} M_s(E)=N_0\frac{\Delta_0}{\sqrt{E^2-\Delta^2}}\]
The density of states, $N_{BCS}(E)$ diverges at $E=\Delta$. 
Near $T_{c}$, where $|f'(\Delta)|$ is still large, the
divergence of $N_{BCS}(E)$
 gives rise to a divergence of $1/T_{1}$. However, the life-time effect
of quasi-particles by the electron-phonon and/or the electron-electron
 interactions, and the anisotropy of the energy gap due to the crystal
 structure broadens the quasi-particle  density of states.  This results in the suppression of the divergence of $1/T_{1}$.
Instead, a peak of $1/T_{1}$ is seen just below $T_{c}$. 
This peak, which is characteristic of singlet pairing, was observed first by Hebel and Slichter\cite{Hebel1959} in the nuclear relaxation rate of Al, and is called the coherence peak.
 At low temperatures, $1/T_{1}$ decreases exponentially due to the uniform gap.

\begin{figure}[htbp]
\centering
\includegraphics[width=8.5cm]{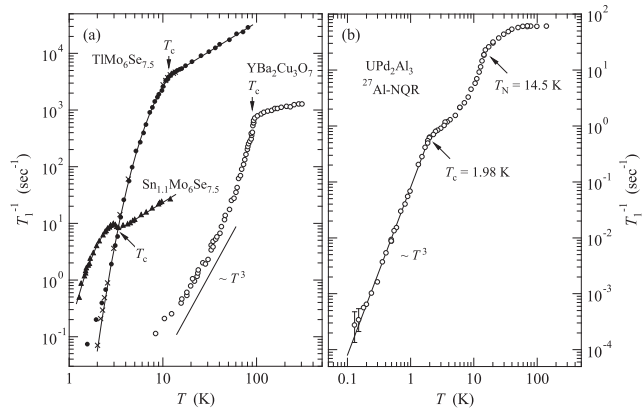}
 \caption[]{\footnotesize (a) Temperature dependences of $1/T_1$ of $^{205}$Tl and
$^{119}$Sn in strong  and weak coupling $s$-wave superconductors,
TlMo$_6$Se$_{7.5}$ and Sn$_{1.1}$Mo$_6$Se$_{7.5}$ and of $^{63}$Cu in
YBa$_2$Cu$_3$O$_7$ at
zero field. Solid lines above and below $T_c$ represent the $T_1T$=constant law and the exponential law with $1/T_1=A\exp{(-\Delta/k_{B}T)}$, respectively. $2\Delta/k_{B}T_{c}$=4.5 and 3.6 are obtained
for TlMo$_6$Se$_8$ and SnMo$_6$Se$_8$, respectively.\cite{Ohsugi1992} (b) Temperature dependence of $^{27}(1/T_1)$ in zero field Al-NQR for UPd$_2$Al$_3$.\cite{Tou2} The solid line shows $T^3$ dependence deduced from the $d$-wave model by using $\Delta(\theta)=\Delta\cos(\theta)$, with $2\Delta/k_{B}T_{c}$=5.5.}
\label{fig3_T1-SC}
\end{figure}

The coherence peak is not always seen in the case of strong-coupling superconductors. As an example of  the relaxation behavior for weak and strong coupling $s$-wave superconductors,
 Fig.\ref{fig3_T1-SC}(a) shows $1/T_{1}$ of $^{119}$Sn and $^{205}$Tl in the Chevrel phase superconductors, Sn$_{1.1}$Mo$_{6}$Se$_{7.5}$ and TlMo$_{6}$Se$_{7.5}$ with $T_{c}$=4.2 K and $T_{c}$=12.2 K, respectively.\cite{Ohsugi1992}
In the normal state, the $T_{1}T$=constant law holds  for both compounds. In the superconducting state, $1/T_{1}$ of $^{119}$Sn in
Sn$_{1.1}$Mo$_{6}$Se$_{7.5}$ has a coherence peak
just below $T_{c}$ and decreases exponentially with $2\Delta=3.6k_{B}T_{c}$, while $1/T_{1}$ of $^{205}$Tl in the strong coupling superconductor TlMo$_{6}$Se$_{7.5}$ has no coherence peak just below $T_{c}$ and decreases exponentially over five orders of magnitude below $0.8T_{c}$ (10 K) with $2\Delta=4.5k_{B}T_{c}$. Even though the coherence peak is depressed, the $s$-wave picture is evidenced by the exponential decrease of $1/T_{1}$ below $T_{c}$.
Figures \ref{fig3_T1-SC}(a) and 3(b) also show another important example, the high-$T_{c}$ cuprates and HF antiferromagnetic superconductor UPd$_2$Al$_3$ \cite{Tou2}. When the gap disappears on a line or points on the Fermi surface for $p$- or $d$-wave pairing, the divergence at the gap edge is weak. Then, note that for the spin-triplet $p$-wave $M_s(E)=0$, and for the spin-singlet $d$-wave $\Delta(\theta,\phi)$ in $M_s(E)$ disappears when averaged out over the Fermi surface. In the case of a line node, $N_s(E)\sim E$ for small $E$ and $1/T_1$ is given by 
\[\frac{1}{T_{1}} \propto \int_{0}^{\infty}E^{2}f(E)(1-f(E))dE \propto
T^{3}.\]

In these compounds listed up in the Table I, the dynamical magnetic properties are different in each case. Nevertheless in all these the superconducting energy gap vanishes along lines on the Fermi surface as evidenced from a rather universal behavior of $1/T_1\propto T^3$ even in the $d$-wave for high-$T_c$ oxides and the $p$-wave for Sr$_2$RuO$_4$.
It is almost certain that anisotropic order parameters with
spin singlet are realized in CeCu$_2$Si$_2$ and
UPd$_2$Al$_3$, and that a different anisotropic order parameter with non-unitary spin-triplet pairing is realized in UPt$_3$ which is the first example of this pairing symmetry in charged many body systems, and UNi$_2$Al$_3$ is reported to belong to a class of spin-triplet pairing as well \cite{Ishida2}.  These variations of the anisotropic order parameter could be due to the different characters of the magnetic fluctuations that lead to the Cooper pairing. We note that the integrated spectral weight of magnetic fluctuations is smaller for Ce compounds with the $4f^1$ than for U compounds with $5f^{2}-5f^{3}$. An important feature is that the ground state adjacent to the antiferromagnetic phase in CeCu$_2$Si$_2$ and CeCu$_2$Ge$_2$ is not the normal phase but the
superconducting one. Next we deal with an intimate interplay between antiferromagnetism and superconductivity.
\section{Novel Superconductivity on the Magnetic Criticality in Heavy-Fermion Systems}
The most common kind of superconductivity (SC) is based on bound electron pairs coupled by deformation of the lattice. However, SC of more subtle origins is rife in strongly correlated electron systems including many HF, cuprate and organic superconductors. In particular, a number of studies on $f$-electron compounds revealed that unconventional SC arises at or close to a quantum critical point (QCP), where magnetic order disappears at low temperature as a function of lattice density via application of hydrostatic pressure ($P$) \cite{Jaccard92,Movshovic96,Mathur98,Hegger00}. These findings suggest that the mechanism forming Cooper pairs can be magnetic in origin. Namely, on the verge of magnetic order, the magnetically soft electron liquid can mediate spin-dependent attractive interactions between the charge carriers \cite{Mathur98}.  
However, the nature of SC and magnetism is still unclear when SC appears very close to the antiferromagnetism (AFM). Therefore, in light of an exotic interplay between these phases, unconventional electronic and magnetic properties around QCP have attracted much attention and a lot of experimental and theoretical works are being extensively made.
\begin{figure}[h]
\centering
\includegraphics[width=8cm]{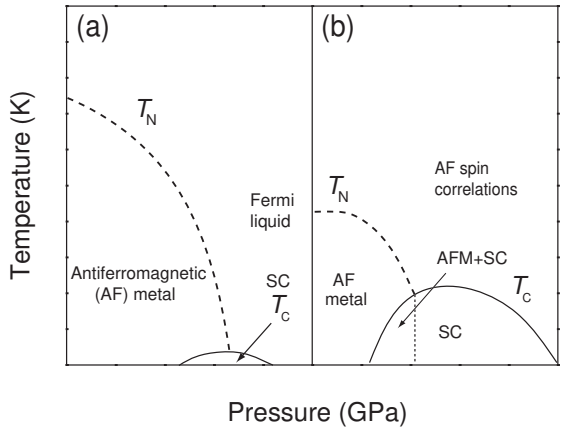}
\caption[]{\footnotesize Schematic phase diagrams of HF compounds: (a) CePd$_2$Si$_2$ \cite{Mathur98,Grosche96,Grosche01}, CeIn$_3$ \cite{Mathur98,Grosche01,Walker97,Muramatsu01,Knebel02} and CeRh$_2$Si$_2$ \cite{Movshovic96,Araki}: (b) CeCu$_2$Si$_2$\cite{Steglich1979,Bellarbi,Bellarbi2,kawasaki01,Ykawasaki2} and CeRhIn$_5$.\cite{Hegger00,Muramatsu} Dotted and solid lines indicate the $P$ dependence of $T_N$ and $T_c$, respectively.}
\label{fig4_P-T_phase}
\end{figure}
The phase diagram, schematically shown in Fig. \ref{fig4_P-T_phase}(a), has been observed in antiferromagnetic HF compounds such as CePd$_2$Si$_2$,\cite{Mathur98,Grosche96,Grosche01} CeIn$_3$,\cite{Mathur98,Grosche01,Walker97,Muramatsu01,Knebel02} and CeRh$_2$Si$_2$. \cite{Movshovic96,Araki} Markedly different behavior, schematically shown in Fig.\ref{fig4_P-T_phase}(b), has been found in the archetypal HF superconductor CeCu$_2$Si$_2$ \cite{Steglich1979,Bellarbi,Bellarbi2,kawasaki01,Ykawasaki2} and the more recently discovered CeRhIn$_5$.\cite{Hegger00,Muramatsu} Although an analogous behavior relevant to a magnetic QCP has been demonstrated in these compounds, it is noteworthy that the associated superconducting region extends to higher densities than in the other compounds; their value of $T_c$ reaches its maximum away from the verge of AFM.\cite{Bellarbi,Bellarbi2,Muramatsu}

In the following sections, we review the recent studies under $P$ on CeCu$_2$Si$_2$, CeRhIn$_5$ and CeIn$_3$ via NQR measurements. These systematic works have revealed the homogeneous coexistent phase of SC and AFM and that its novel superconducting nature exhibits the gapless nature in the low-lying excitations below $T_c$, which differ from the superconducting characteristics for the HF superconductors reported to possess the line-node gap.\cite{Ykawasaki2,Mito01,Shinji03} 
\section{Exotic Magnetism and Superconductivity on the Magnetic Criticality in CeCu$_2$Si$_2$}
\subsection{The temperature versus pressure phase diagram}
\begin{figure}[tbp]
\centering
\includegraphics[width=7.5cm]{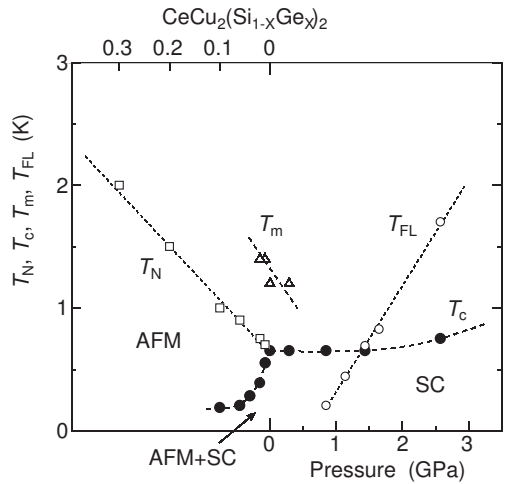}
\caption[]{\footnotesize The combined phase diagram of AFM and SC for CeCu$_2$(Si$_{1-x}$Ge$_{x}$)$_2$ and for Ce$_{0.99}$Cu$_{2.02}$Si$_2$ under $P$.
$T_N$ and $T_c$ are the respective transition temperature of AFM and SC\@. Also shown are $T_m$ below which antiferromagnetic critical fluctuations develop and $T_{FL}$ below which $1/T_1T$ becomes constant.
}
\label{fig5_CeCuSi}
\end{figure}

The firstly-discovered HF superconductor CeCu$_2$Si$_2$ is located just at the border to the AFM at $P=0$.\cite{Gegenwart98,Ishida1}
This was evidenced by various magnetic anomalies observed above $T_c$ \cite{nakamura} and by the fact that the magnetic {\it A-phase} appears when SC is suppressed by a magnetic field $H$.\cite{Bruls94}
Furthermore, the transport, thermodynamic and NQR measurements  consistently indicated that nominally off-tuned Ce$_{0.99}$Cu$_{2.02}$Si$_2$ is located just at $P_c$ and crosses its QCP by applying a minute pressure of $P=0.2$ GPa.\cite{Gegenwart98,kawasaki01} The magnetic and superconducting properties in CeCu$_2$Si$_2$ were investigated around the QCP as the functions of $P$ for Ce$_{0.99}$Cu$_{2.02}$Si$_2$ just at the border to AFM and of Ge content $x$ for CeCu$_2$(Si$_{1-x}$Ge$_x$)$_2$ by Cu-NQR measurements \cite{kawasaki01,kawasaki02}. Figure \ref{fig5_CeCuSi} shows the phase diagram referred from the literature.\cite{kawasaki02}
Here, $T_{FL}$ is an effective Fermi temperature below which $1/T_1 $ divided by temperature ($1/T_1T$) stays constant and $T_m$ is a temperature below which antiferromagnetic critical fluctuations  start to develop.
Note that a primary effect of Ge doping expands the lattice \cite{trovarelli97} and that its chemical pressure is $-0.076$ GPa per 1\% Ge doping as suggested from the $P$ variation of Cu-NQR frequency $\nu_Q$ in CeCu$_2$Ge$_2$ and CeCu$_2$Si$_2$.\cite{kitaoka95}
In the normal state, the antiferromagnetic critical fluctuations develop below $T_m \sim 1.2$ K without the onset of AFM. The exotic SC emerges in Ce$_{0.99}$Cu$_{2.02}$Si$_2$ below $T_c \sim 0.65$ K, where antiferromagnetic critical fluctuations remain active even below $T_c$. 
\subsection{The uniformly coexistent phase of AFM and SC in CeCu$_2$(Si$_{1-x}$Ge$_x$)$_2$}
\begin{figure}[htbp]
\centering
\includegraphics[width=7cm]{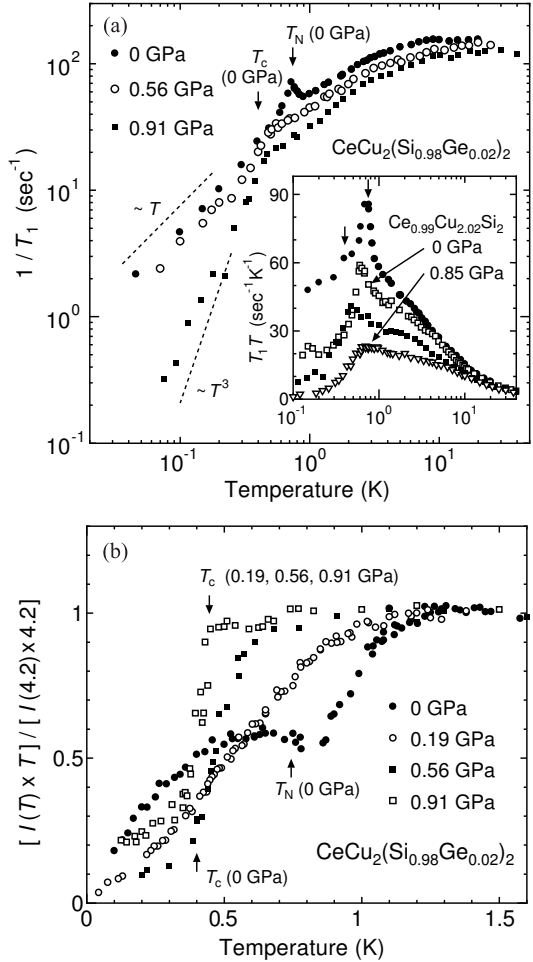}
\caption[]{\footnotesize (a) The $T$ dependence of $1/T_1$ of CeCu$_2$(Si$_{0.98}$Ge$_{0.02}$)$_2$ at several pressures.
Inset shows the $T$ dependence of $1/T_1T$ of CeCu$_2$(Si$_{0.98}$Ge$_{0.02}$)$_2$ at $P$ = 0 GPa (closed circles) and 0.91 GPa (closed squares) and those of Ce$_{0.99}$Cu$_{2.02}$Si$_2$ at $P$ = 0 (open squares) and 0.85 GPa (open triangles).
Arrows indicate $T_N \sim 0.75$ K and $T_c \sim$ 0.4 K at $P = 0$ GPa for CeCu$_2$(Si$_{0.98}$Ge$_{0.02}$)$_2$\@. (b) The $T$ dependence of $I(T)\times T$ at several pressures, where $I(T)$ is an NQR intensity normalized by the value at 4.2 K\@.}
\label{fig6_CeCuSi_T1+int}
\end{figure}

Markedly by substituting only 1\% Ge, AFM emerges at $T_N \sim$ 0.7 K, followed by the SC at $T_c \sim$ 0.5 K. Unexpectedly, $1/T_1$ does not show any significant reduction at $T_c$, but follows a $1/T_1T$ = const.\ behavior well below $T_c$ as observed in Ce$_{0.99}$Cu$_{2.02}$Si$_2$\@ as presented in Fig. \ref{fig6_CeCuSi_T1+int}(a). As Ge content increases, $T_N$ progressively increases, while $T_c$ steeply decreases. For the samples at more than $x=0.06$, the magnetic properties above $T_N$ progressively change to those in a localized regime as observed in CeCu$_2$Ge$_2$.\cite{kitaoka95}

Further insight into the exotic SC is obtained on CeCu$_2$(Si$_{0.98}$Ge$_{0.02}$)$_2$ that reveals the uniformly coexistent phase of AFM ($T_N \sim$ 0.75 K) and SC ($T_c \sim$ 0.4 K) under $P = 0$. 
 Figure \ref{fig6_CeCuSi_T1+int}(a) shows the $T$ dependence of $1/T_1$ at $P$ = 0 GPa (closed circles), 0.56 GPa (open circles) and 0.91 GPa (closed squares). In the entire $T$ range, $1/T_1$ is suppressed with increasing $P$. Once AFM is suppressed at pressures exceeding $P = 0.19$ GPa, any trace of anomaly associated with it is not observed at all down to $T_c \sim$ 0.45 K at $P$ = 0.56 GPa and 0.91 GPa\@.
It is, therefore, considered that the AFM in CeCu$_2$(Si$_{0.98}$Ge$_{0.02}$)$_2$ is not triggered by some disorder effect but by the intrinsic lattice expansion due to the Ge doping.

In order to demonstrate a systematic evolution of antiferromagnetic critical fluctuations at the paramagnetic state, the inset of Fig. \ref{fig6_CeCuSi_T1+int}(a) shows the $T$ dependence of $1/T_1T$ in CeCu$_2$(Si$_{0.98}$Ge$_{0.02}$)$_2$ at $P$ = 0 GPa (closed circles) and 0.91 GPa (closed squares) along with the results in Ce$_{0.99}$Cu$_{2.02}$Si$_2$ at $P$ = 0 GPa (open squares) and 0.85 GPa (open triangles).\cite{kawasaki01,kawasaki02}
The result of $1/T_1T$ in CeCu$_2$(Si$_{0.98}$Ge$_{0.02}$)$_2$ at $P = 0$ GPa is well explained by the spin-fluctuations theory for weakly itinerant AFM in $T_c < T < 1.5$ K.\cite{kitaoka01,kawasaki01,kawasaki02}
The good agreement between the experiment and the calculation indicates that a long-range nature of the AFM is in the itinerant regime.
At $P = 0.91$ GPa, $1/T_1T$ is suppressed and resembles a behavior that would be expected at an intermediate pressure between $P = 0$ GPa and 0.85 GPa for Ce$_{0.99}$Cu$_{2.02}$Si$_2$\@.

Next, we discuss an intimate $P$-induced evolution of antiferromagnetic critical fluctuations in the superconducting state.
As seen in Fig. \ref{fig6_CeCuSi_T1+int}(a) and its inset, the $1/T_1$ and $1/T_1T$ at $P =$ 0 GPa do not show a distinct reduction below $T_c$, but instead, a $T_1T=$ const.\ behavior emerges well below $T_c$.
At $P$ = 0.56 GPa, the AFM is depressed, but antiferromagnetic critical fluctuations develop in the normal state. It is noteworthy that the relation of $1/T_1\propto T$ is still valid below $T_c$, resembling the behavior for Ce$_{0.99}$Cu$_{2.02}$Si$_2$ at $P = 0$ GPa.\cite{kawasaki01,Ishida1}
By contrast, $1/T_1$ at $P = 0.91$ GPa follows a relation of $1/T_1 \propto T^3$ below $T_c \sim$ 0.45 K, consistent with the line-node gap.
This typical HF-SC was observed in Ce$_{0.99}$Cu$_{2.02}$Si$_2$ at pressures exceeding $P=$ 0.85 GPa as well.\cite{kawasaki01}
Therefore, it is considered that the unconventional SC at $P = 0$ GPa and 0.56 GPa evolves into the typical HF-SC with the line-node gap at pressures exceeding $P = $0.91 GPa\@.
Apparently, these results exclude a possible impurity effect as a primary cause for the $T_1T$ = const.\ behavior below $T_c$ at $P =$ 0 GPa\@.
We point out that a reason why the $1/T_1$ at $P = 0$ GPa is deviated from $1/T_1\propto T^3$ below $T_c$ is ascribed not to the impurity effect but to the presence of low-lying excitations inherent to the exotic superconducting state.
\subsection{Evidence for antiferromagnetic critical fluctuations at the boundary between AFM and SC}
Figure \ref{fig6_CeCuSi_T1+int}(b) indicates the $T$ dependence of NQR intensity multiplied by temperature $I(T)\times T$ in CeCu$_2$(Si$_{0.98}$Ge$_{0.02}$)$_2$ at $P =$ 0, 0.19, 0.56, and 0.91 GPa\@.
Here, the $I(T)$ normalized by the value at 4.2 K is an integrated intensity over frequencies where NQR spectrum was observed.
Note that $I(T)\times T$ stays constant generally, if $T_1$ and/or $T_2$ range in the observable time window that is typically more than several microseconds.
Therefore, the distinct reduction in $I(T)\times T$ upon cooling is ascribed to the development of antiferromagnetic critical fluctuations, since it leads to an extraordinary short relaxation time of $\sim$ 0.14 $\mu$sec.\cite{Ishida1}

The $I(T)\times T=1$ at $P = 0$ GPa decreases down to about $I(T)\times T=0.55$ at $T_N \sim 0.75$ K upon cooling below $T_m \sim$ 1.2 K\@.
Its reduction stops around $T_N$, but does no longer recover with further decreasing $T$\@.
Note that its reduction below $T_c \sim 0.4$ K is due to the superconducting diamagnetic shielding of rf field for the NQR measurement.
As $P$ increases, $T_m$ becomes smaller, in agreement with the result presented in the phase diagram of Fig. \ref{fig5_CeCuSi}, and the reduction in $I(T)\times T$  becomes moderate in the normal state.
With further increasing $P$ up to 0.91 GPa, eventually, $I(T)\times T$ remains nearly constant down to $T_c \sim 0.45$ K, indicative of no anomaly related to antiferromagnetic critical fluctuations.
This behavior resembles the result observed at pressures exceeding 
$P = 0.85$ GPa in Ce$_{0.99}$Cu$_{2.02}$Si$_2$\@.
These results also assure that the Ge substitution expands the lattice of Ce$_{0.99}$Cu$_{2.02}$Si$_2$\@.
It seems, therefore, that these exotic SC could be rather robust against the persistence of antiferromagnetic critical fluctuations and yet the appearance of AFM.
\subsection{Towards a new concept for superconductivity}
Antiferromagnetic critical fluctuations develop below $T_{\rm m}$ in $0 \leq x < 0.06$ in CeCu$_2$(Si$_{1-x}$Ge$_x$)$_2$ and $0 < P < 0.2$ GPa in CeCu$_2$Si$_2$ \@. {\it Remarkably this antiferromagnetic critical fluctuations, which emerge closely at the border between AFM and SC, is not in a completely static regime, but in a dynamical one with a characteristic frequency $\omega_c\sim$ 3 MHz}.\cite{koda} A $T_1\sim$ 0.14 $\mu$sec, that is obtained from the relation $1/T_1 \propto \omega_c(T)/(\omega_c(T)^2+\omega_{NQR}^2)$, are too fast for the NQR signal to be observed.\cite{Ishida1} Here, $\omega_{NQR}$ ($\sim$ 3.4 MHz) is a central NQR frequency. Neutron scattering experiments have, on the other hand, observed the long-range incommensurate AFM as the nature of antiferromagnetic critical fluctuations \cite{Stockert} because its resolution of characteristic frequencies is higher than in the NQR experiments. As a matter of fact,
once  a small amount of Ge are substituted for Si to expand its lattice, antiferromagnetic critical fluctuations are suddenly replaced by the static AFM,\cite{trovarelli97,kawasaki02} whereas they survive down to 0.012 K at $x=0$.
With increasing $x$, $T_{\rm N}$ progressively increases, while $T_{\rm c}$ steeply decreases. Correspondingly, antiferromagnetic critical fluctuations are suppressed for the samples at more than $x$ = 0.06.
It is noteworthy that the static AFM seems to suddenly disappear at $x$ = 0 as if $T_{\rm N}$ were replaced by $T_{\rm c}$\@.
Eventually, the SC at $x = 0$ emerges under the dominance of antiferromagnetic critical fluctuations.\cite{koda} 
However, when the application of $H$ suppresses the SC, the first-order like transition from SC to a magnetic {\it A-phase} takes place.\cite{Bruls94} Since $T_N$ becomes comparable to $T_c$, some superconducting fluctuations may prevent the onset of static AFM.\cite{kitaoka01}
Once the application of pressure exceeding $P=$0.2 GPa suppresses antiferromagnetic critical fluctuations, the typical HF-SC takes place with the line-node gap.

In CeCu$_2$(Si$_{1-x}$Ge$_x$)$_2$, one $4f$ electron per Ce ion plays vital role for both SC and AFM, leading to the novel states of matter. We have proposed that the uniformly coexistent phase of AFM and SC in the slightly Ge substituted compounds, and the magnetic-field induced {\it A-phase} for the homogeneous CeCu$_2$Si$_2$ in $0 < P < 0.2$ GPa are accounted for on the basis of an SO(5) theory.\cite{kitaoka01,zhang97} 
Concerning the interplay between AFM and SC, we would propose that the antiferromagnetic critical fluctuations in the superconducting state  at $x$ = 0 may correspond to a pseudo Goldstone mode due to the broken U(1) symmetry. Due to the closeness to the magnetic QCP, however, such the gapped mode in the SC should be characterized by an extremely tiny excitation (resonance) energy. 

The intimate interplay between SC and AFM found in 
CeCu$_2$Si$_2$ has been a long-standing problem - unresolved for over a decade. We have proposed that the SO(5) theory constructed on the basis of quantum-field theory may give a coherent interpretation for these exotic phases found in CeCu$_2$Si$_2$.\cite{kitaoka01} In this context, we would suggest that the SC in CeCu$_2$Si$_2$ could be mediated by the {\it same magnetic interaction} as leads to the AFM in CeCu$_2$(Si$_{1-x}$Ge$_x$)$_2$. This is in marked contrast to the BCS superconductors in which the pair binding is mediated by phonons $-$ vibrations of the lattice density.
\section{Uniformly Coexistent Phase of Antiferromagnetism and Superconductivity in CeRhIn$_5$ under Pressure}
\subsection{The temperature versus pressure phase diagram}
\begin{figure}[htbp]
\centering
\includegraphics[width=7cm]{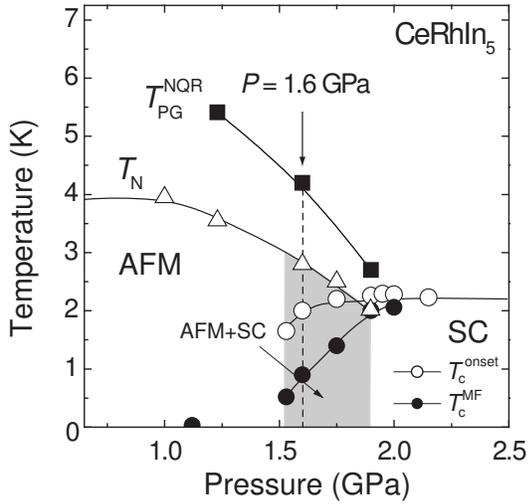}
\caption[]{\footnotesize The $P$ vs $T$ phase diagram for CeRhIn$_5$.
The respective marks denoted by solid square and open triangle  correspond to the pseudogap temperature $T_{PG}^{NQR}$ and the antiferromagnetic ordering temperature $T_N$ at the In site. The open and solid circles correspond to the onset temperature $T_c^{onset}$ and $T_c^{MF}$ of superconducting transition (see text). Dotted line denotes the position for $P=1.6$ GPa. Shaded region indicates the $P$ region where both phases of AFM and SC coexist uniformly.}
\label{fig7_CeRhIn}
\end{figure}

A new antiferromagnetic HF compound CeRhIn$_5$ undergoes the helical magnetic order at a N\'eel temperature $T_N=3.8$ K with an incommensurate wave vector ${\rm q_M=(1/2, 1/2, 0.297)}$.\cite{Curro} A neutron experiment revealed the reduced Ce magnetic moments $M_s\sim$ 0.8$\mu_{\rm B}$.\cite{christianson,Bao} The $P$-induced transition from AFM to SC takes place at a relatively lower critical pressure $P_c=1.63$ GPa and higher $T_c = 2.2$ K than in previous examples.\cite{Jaccard92,Mathur98,Grosche96,Movshovic96,Hegger00,Walker97}
Figure \ref{fig7_CeRhIn} indicates the $P$ vs $T$ phase diagram of CeRhIn$_5$ for AFM and SC that was determined by the In-NQR measurements under $P$.  
The NQR study showed that $T_N$ gradually increases up to 4 K as $P$ increases up to $P = 1.0$ GPa and decreases with further increasing $P$. \cite{Mito01,Mito03,Shinji} In addition, the $T$ dependence of $1/T_1$ probed the pseudogap behavior at $P = 1.23$ and 1.6 GPa.\cite{Shinji} This suggests that CeRhIn$_5$ may resemble other strongly correlated electron systems.\cite{Timusk,Kanoda} Note that the value of bulk superconducting transition temperature $T_c^{MF}$ is progressively reduced as shown by closed circle in Fig. \ref{fig7_CeRhIn}. For $P_c= 2.0$ GPa $< P $  where AFM disappears, {\it $1/T_1$ decreases obeying a $T^3$ law} without the coherence peak just below $T_c$.\cite{Mito01,Yashima} This indicates that the SC of CeRhIn$_5$ is unconventional with the line-node gap.\cite{Mito01,KohoriEPJ00}  
\begin{figure}[htbp]
\centering
\includegraphics[width=7.5cm]{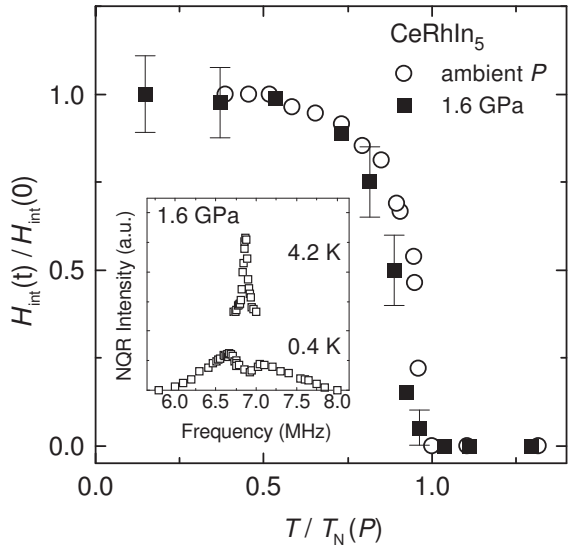}
\caption[]{\footnotesize Plots of  $H_{int}(T)/H_{int}(0)$ vs $T/T_N$ at $P$ = 0 and 1.6 GPa (see text). Inset shows the $^{115}$In-NQR spectra of 1$\nu_Q$ at $P$ = 1.6 GPa above and below $T_N$ = 2.8 K. }
\label{fig8_CeRhIn_Spec}
\end{figure}
\subsection{Gapless magnetic and quasi-particles excitations due to the uniformly coexistent phase of AFM and SC}
We present microscopic evidence for the exotic SC at the uniformly coexistent phase of AFM and SC in CeRhIn$_5$ at 
$P = 1.6$ GPa. The inset of Fig. \ref{fig8_CeRhIn_Spec} displays the NQR spectra above and below $T_N$ at $P = 1.6$ GPa. Below $T_N = 2.8$ K, the NQR spectrum splits into two peaks due to the appearance of $H_{int}$ at the In site. This is clear evidence for the occurrence of AFM at $P = 1.6$ GPa. The plots of $H_{int}(T)/H_{int}(0)=M_s(T)/M_s(0)$ vs $(T/T_N)$ at $P = 0$ and 1.6 GPa are compared in Fig. \ref{fig8_CeRhIn_Spec}, showing nearly the same behavior. Here $H_{int}(0)$ is the value extrapolated to zero at $T = 0$ K and $M_s(T)$ is the $T$ dependence of spontaneous staggered magnetic moment. The character of AFM at $P = 1.6$ GPa is expected to be not so much different from that at $P = 0$. 

\begin{figure}[htbp]
\centering
\includegraphics[width=8cm]{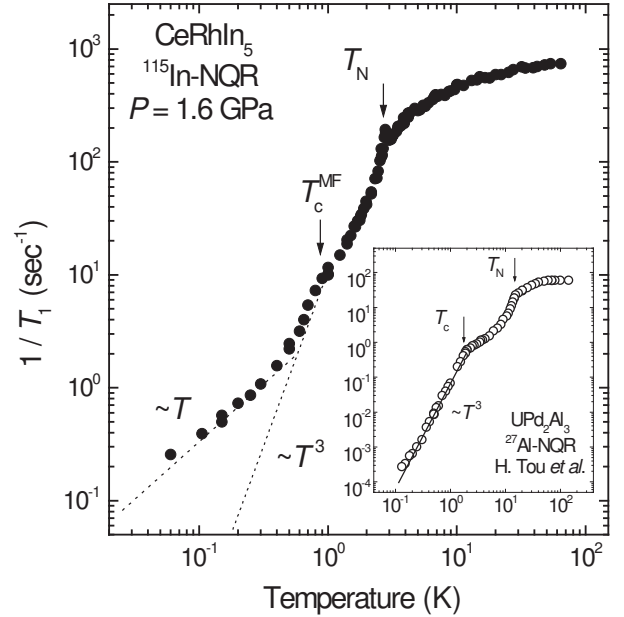}
\caption[]{\footnotesize The $T$ dependence of $1/T_1$ at $P$ = 1.6 GPa.  Both dotted lines correspond to $1/T_1\propto T$ and $1/T_1\propto T^3$.  Inset indicates the $T$ dependence of $^{27}$Al-NQR $1/T_1$ of UPd$_2$Al$_3$ cited from the literature.\cite{Tou2} Solid line corresponds to $1/T_1\propto T^3$.}
\label{fig9_CeRhIn_T1}
\end{figure}

Figure \ref{fig9_CeRhIn_T1} indicates the $T$ dependence of $1/T_1$ at $P = 1.6$ GPa. A clear peak in $1/T_1$ is due to  antiferromagnetic critical fluctuations at $T_N = 2.8$ K. Below $T_N = 2.8$ K, $1/T_1$ continues to decrease moderately down to $T_c^{MF} = 0.9$ K even though passing across $T_c^{onset}\sim 2$ K. This relaxation behavior suggests that SC does not develop following the mean-field (MF) approximation below $T_c^{onset}$.  Markedly, $1/T_1$ decreases below $T_c^{MF}$, exhibiting a faint $T^3$ behavior in a narrow $T$ range. With further decreasing $T$, $1/T_1$ becomes proportional to the temperature, indicative of the gapless nature in low-lying excitation spectrum in the uniformly coexistent state of SC and AFM. Thus the $T_1$ measurement has revealed that the intimate interplay between AFM and SC  gives rise to an {\it amplitude fluctuation of superconducting order parameter}  between $T_c^{onset}$ and $T_c^{MF}$. 
Such fluctuations may be responsible for the broad transition in resistance and ac-susceptibility ($\chi_{ac}$) measurements. Furthermore, the $T_1T=$ const. behavior well below $T_c^{MF}$ evidences the gapless nature in low-lying excitations at the uniformly coexistent phase of AFM and SC. This result is consistent with those in CeCu$_2$Si$_2$ at the border to AFM \cite{kawasaki01} and a series of CeCu$_2$(Si$_{1-x}$Ge$_2$)$_2$ compounds that show the uniformly coexistent phase of AFM and SC.\cite{kitaoka01,kawasaki02}  
It is noteworthy that such the $T_1T$ = const. behavior is not observed  below $T_c$ at $P$ = 2.1 GPa.\cite{Mito01} This means that the origin for the $T_1T=$ const. behavior below $T_c^{MF}$ at $P = 1.6$ GPa is not associated with some impurity effect. If it were the case, the residual density of states below $T_c$ should not depend on $P$. 
This novel feature differs from the uranium(U)-based HF antiferromagnetic superconductor UPd$_2$Al$_3$ which has multiple 5$f$ electrons. In UPd$_2$Al$_3$, a superconducting transition occurs at $T_c$ = 1.8 K well below $T_N$ = 14.3 K.\cite{Geibel,Steglich2} As indicated in Fig.\ref{fig3_T1-SC}(b) and the inset of Fig. \ref{fig9_CeRhIn_T1},\cite{Tou2} in UPd$_2$Al$_3$, {\it $1/T_1$ decreases obeying a $T^3$ law over three orders of magnitude} below the onset of $T_c$ without any trace for the $T_1T$ = const. behavior. This is consistent with the line-node gap even in the uniformly coexistent phase of AFM and SC. 
\subsection{Superconducting fluctuations due to the uniformly coexistent phase of AFM and SC} 
\begin{figure}[htbp]
\centering
\includegraphics[width=8cm]{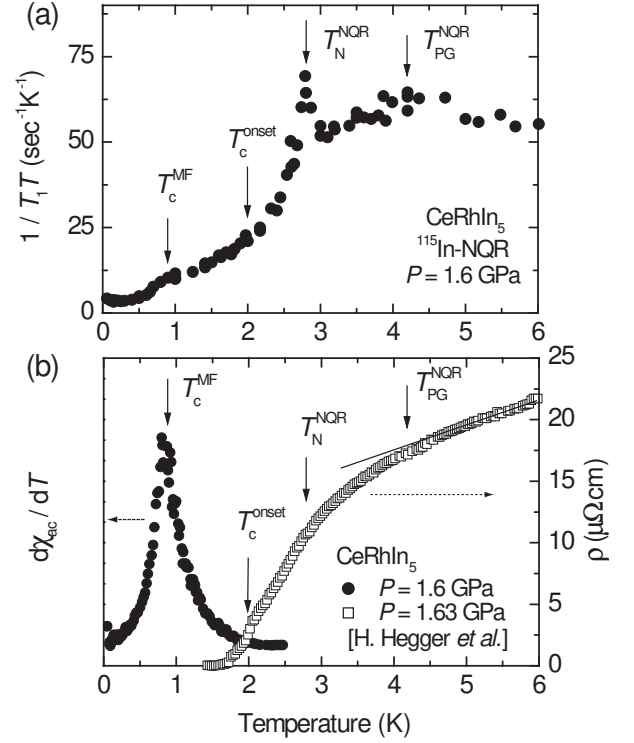}
\caption[]{\footnotesize (a) The $T$ dependence of $1/T_1T$ at $P = 1.6$ GPa. (b) The $T$ dependencies of $d\chi_{ac}/dT$ at $P = 1.6$ GPa and resistance at 
$P = 1.63$ GPa cited  from the literature.\cite{Hegger00} $T_c^{MF}$ and $T_c^{onset}$ correspond to the respective temperatures at which $d\chi_{ac}/dT$ has a peak and below which $\chi_{ac}$ starts to decrease. $T_N$ corresponds to the antiferromagnetic ordering temperature at which $1/T_1T$ exhibits a peak and $T_{PG}^{NQR}$ to the pseudogap temperature below which it starts to decrease. A solid line is an eye guide for the $T$- linear variation in resistance at temperatures higher than $T_{PG}^{NQR}$.}
\label{fig10_CeRhIn_T1T}
\end{figure}

In order to highlight the novel SC on a microscopic level, the $T$ dependence of $1/T_1T$ is shown in Fig. \ref{fig10_CeRhIn_T1T}(a) at $P = 1.6$ GPa in $T = 0.05 - 6$ K and is compared with the $T$ dependence of the resistance $R(T)$ at $P = 1.63$ GPa referred from the literature.\cite{Hegger00} Although each value of $P$ is not exactly the same, they only differ by 2\%.  We remark that the $T$ dependence of $1/T_1T$ points to the pseudogap behavior around $T_{PG}^{NQR}$ = 4.2 K, the AFM at $T_N$ = 2.8 K, and the SC at $T_c^{MF}$ = 0.9 K at which $d\chi_{ac}/dT$ has a peak as seen in Fig. \ref{fig10_CeRhIn_T1T}(b). This result itself evidences the uniformly coexistent state of AFM and SC.  A comparison of $1/T_1T$ with  the $R(T)$ at $P = 1.63$ GPa in Fig. \ref{fig10_CeRhIn_T1T}(b) is informative in shedding light on the uniqueness of AFM and SC. Below $T_{PG}^{NQR}$, $R(T)$ starts to decrease more rapidly than a $T$-linear variation extrapolated from a high $T$ side. It continues to decrease across $T_N$ = 2.8 K, reaching the zero resistance at $T_c^{zero}\sim$ 1.5 K.

The resistive superconducting transition width becomes broader. Note that this broad transition is not due to the distribution in $T_c$ caused by some distribution in values of pressure because the value of $T_1$, the microscopic quantity, is uniquely determined in the $T$ range in concern. Unexpectedly, $T_c^{onset}\sim$ 2 K, that is defined as the temperature below which the diamagnetism starts to appear, is higher than $T_c^{zero}\sim$ 1.5 K. Any signature for the onset of SC from the $1/T_1$ measurement is not evident in between $T_c^{onset}$ and $T_c^{MF}$, demonstrating that the mean-field type of gap does not grow up down to $T_c^{MF}\sim$ 0.9 K.  In $T_c^{onset}> T > T_c^{MF}$, therefore, {\it the SC is not in the conventional mean-field regime that accompanies a gap formation at the Fermi level below $T_c^{onset}$, but in a gapless regime with a pairing correlation.} In this context, we would propose that the presence of some fluctuations of antiferromagnetic order parameter (OP) or magnetic density fluctuations bring about the fluctuations of superconducting OP, making the superconducting transition very broad.  
As the antiferromagnetic OP fully develops at temperatures lower than $T_c^{onset}$, the SC reveals the gap formation below $T_c^{MF}$=0.5$T_c^{onset}$.    
\subsection{A novel interplay between AFM and SC}
Recent neutron-diffraction experiment suggests that the size of staggered moment $M_s$ in the AFM is almost independent of $P$.\cite{Bao} Its relatively large size of moment with $M_s\sim 0.8\mu_B$ seems to support such a picture that the {\it same $f$-electron} exhibits simultaneously itinerant and localized dual nature, because there is only one $4f$-electron per Ce$^{3+}$ ion. In this context, it is natural to consider that the superconducting nature in the uniformly coexistent phase of AFM and SC belongs to a novel class of phase which differs from the unconventional $d$-wave SC with the line-node gap. As a matter of fact, a theoretical model has been recently put forth to address the underlying issue in the uniformly coexistent phase of AFM and SC.\cite{Fuseya}
\section{Emergent Phases of Superconductivity and Antiferromagnetism on the Magnetic Criticality in CeIn$_3$}
\subsection{The temperature versus pressure phase diagram}
\begin{figure}[h]
\centering
\includegraphics[width=7cm]{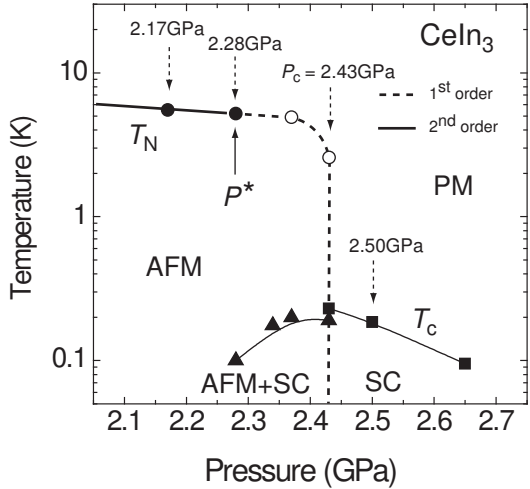}
\caption[]{\footnotesize The $P$ vs $T$ phase diagram of CeIn$_3$ determined by the In-NQR measurement under $P$. Solid circle denotes the N\'{e}el temperature $T_N$ where antiferromagnetic order takes place. $P^*=2.28$ GPa is an end point of the first-order transition and $P_c\sim 2.43$ GPa is the critical pressure at which the first-order transition occurs from AFM to PM. Note that the phase separation into AFM and PM occurs at $P_c$, revealing almost equivalent fraction of each phase. Regardless of this phase separation, SC emerges at both phases.}
\label{fig11_CeIn3}
\end{figure}

Figure \ref{fig11_CeIn3} indicates the $P$ vs $T$ phase diagram in CeIn$_3$ around $P_c$ that is the critical pressure at which a first-order transition occurs from AFM to paramagnetism (PM). This work has deepened the understanding of the physical properties on the verge of AFM in CeIn$_3$ that exhibits the archetypal phase diagram shown in Fig. \ref{fig4_P-T_phase}(a).\cite{Shinji04} At $P = 2.65$ GPa larger than $P_c$, the measurements of $1/T_1$ and $\chi_{ac}$ down to $T = 50$ mK provided the first evidence of unconventional SC at $T_c = 95$ mK in CeIn$_3$, which arises in the HF state fully established below $T_{FL} = 5$ K.\cite{Shinji02In3}
By contrast, in $P = 2.28 - 2.50$ GPa, the phase separation into AFM and PM is evidenced in CeIn$_3$ from the observation of two kinds of NQR spectra. Nevertheless, it is highlighted that the SC in CeIn$_3$ occurs in both AFM and PM at $P_c = 2.43$ GPa. The maximum value of $T_c^{max}=230$ mK is observed for the SC for PM at $P_c$, whereas, markedly, the SC also sets in at $T_c = 190$ mK for AFM with $T_N=2.5$ K. The present results indicate that the first-order phase transition occurs at $P_c\sim 2.43$ GPa from the uniformly coexistent phase of SC and AFM to the single phase of SC under the paramagnetic HF state. Therefore, a QCP is absent in CeIn$_3$.\cite{Shinji04}

CeIn$_3$ forms in the cubic AuCu$_3$ structure and orders antiferromagnetically below $T_N = 10.2$ K at $P = 0$ with an ordering vector {\bf Q} = (1/2,1/2,1/2) and Ce magnetic moment $M_S\sim 0.5\mu_B$, which were determined by NQR measurements \cite{Kohori99,Kohori00} and the neutron-diffraction experiment on single crystals,\cite{Knafo} respectively. The resistivity measurements of CeIn$_3$ have clarified the $P$ vs $T$ phase diagram of AFM and SC: $T_N$ decreases with increasing $P$. On the verge of AFM, the SC emerges in a narrow $P$ range of about 0.5 GPa, exhibiting a maximum value of $T_c\sim 0.2$ K at $P_c = 2.5$ GPa where AFM disappears.\cite{Walker97,Mathur98,Grosche01,Muramatsu01,Knebel02}
\begin{figure}[h]
\centering
\includegraphics[width=7.5cm]{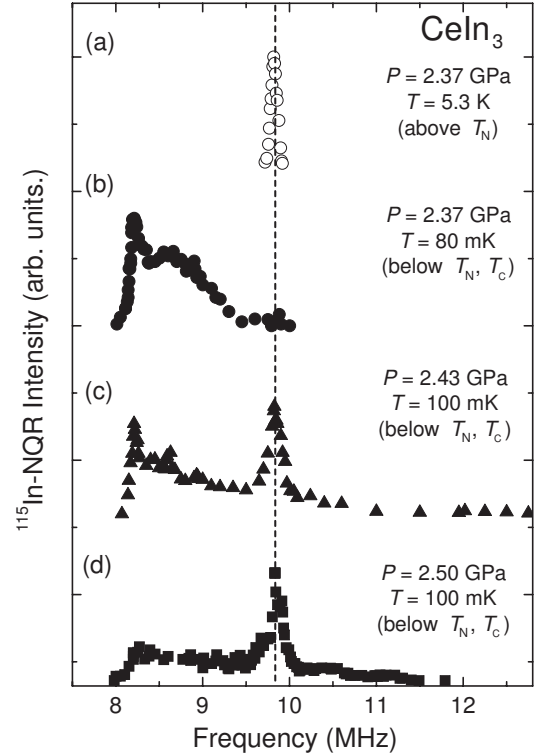}
\caption[]{\footnotesize The $P$ dependence of $^{115}$In NQR spectrum for CeIn$_3$ at (a): $P=2.37$ GPa above $T_N$, and (b): $P=2.37$ GPa, (c): $P=2.43$ GPa and (d): $P=2.50$ GPa at temperatures lower than the $T_N$ and $T_c$. The dotted line indicates the peak position at which the NQR spectrum is observed for PM.}
\label{fig12_CeIn3_Spec}
\end{figure}
\subsection{Evidence for the first-order transition from AFM to PM}
Figure \ref{fig12_CeIn3_Spec} shows the NQR spectra of 1$\nu_{Q}$ transition for the PM at (a) $P = 2.37$ GPa and for temperatures lower than $T_N$ and $T_c$ at (b) $P = 2.37$ GPa, (c) $P_c = 2.43$ GPa and (d) $P = 2.50$ GPa. Note that the 1$\nu_Q$ transition can sensitively probe the appearance of internal field $H_{int}(T)$ associated with even tiny Ce ordered moments on the verge of AFM. As a matter of fact, as seen in Figs. \ref{fig12_CeIn3_Spec}(a) and \ref{fig12_CeIn3_Spec}(b), a drastic change in the NQR spectral shape is observed due to the occurrence of $H_{int}(T)$ at the In nuclei below $T_N$. By contrast, the spectra at Fig.\ref{fig12_CeIn3_Spec}(c) at $P = 2.43$ GPa and Fig.\ref{fig12_CeIn3_Spec}(d) at $P = 2.50$ GPa include two kinds of spectra arising from AFM and PM provides firm microscopic evidence for the emergence of magnetic phase separation.  It should be noted that the phase separations at $P^* = 2.28$ and 2.37 GPa are observed only in the respective ranges $T = 3$ K and $T_N = 5.2$ K and $T = 1$ K and $T_N = 4.9$ K.
\subsection{The novel SC under both the backgrounds of AFM and PM}
\begin{figure}[htbp]
\centering
\includegraphics[width=9cm]{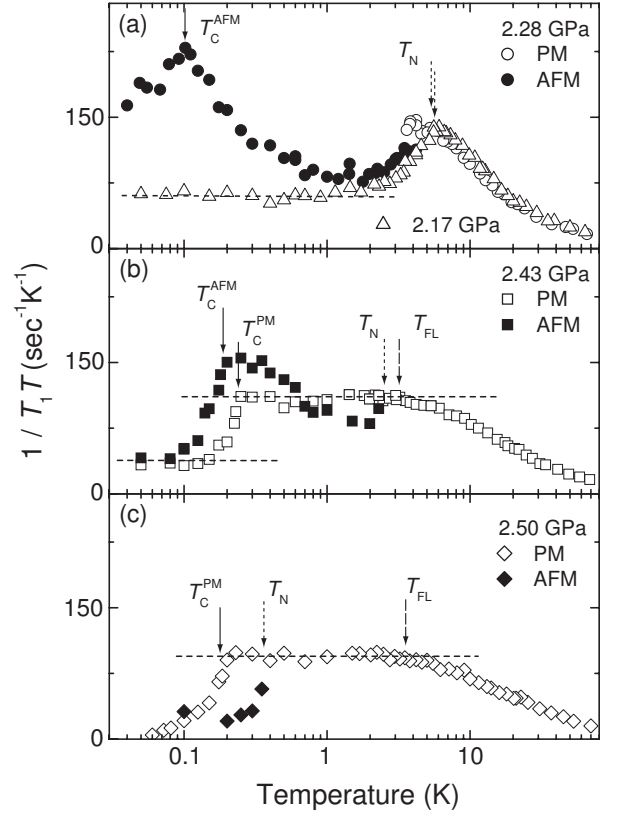}
\caption[]{\footnotesize $T$ dependence of $ ^{115}(1/T_1T)$ in CeIn$_3$ at (a): $P=2.17$ GPa and $P^*=2.28$ GPa, (b): $P_c=2.43$ GPa and (c): $P=2.50$ GPa. Open and solid symbols indicate the data for PM and AFM measured at $\sim$ 9.8 MHz and $\sim$ 8.2 MHz, respectively. The solid arrow indicates the respective superconducting transition temperatures $T_c^{PM}$ and $T_c^{AFM}$ for PM and AFM.  The dotted and dashed arrows indicate, respectively, the $T_N$ and the characteristic temperature $T_{FL}$ below which the $T_1T = $const. law (dotted line) is valid, which is characteristic of the Fermi-liquid state.}
\label{fig13_T1T}
\end{figure}

The uniformly coexistent phase of AFM and SC is corroborated by direct evidence from the $T$ dependence of $1/T_1T$ that can probe the low-lying excitations due to quasiparticles in SC and the magnetic excitations in AFM.  Figure \ref{fig13_T1T} shows the drastic evolution in the $P$ and $T$ dependencies of $1/T_1T$ for AFM (solid symbols) and PM (open symbols) at $P$ = 2.17, 2.28, 2.43 and 2.50 GPa. Here, $T_N$ was determined as the temperature below which the NQR intensity for PM decreases due to the emergence of AFM associated with the magnetic phase separation. The $T_1$ for AFM and PM is separately measured at respective NQR peaks which are clearly distinguished from each other, as shown in Figs.\ref{fig12_CeIn3_Spec}(b), \ref{fig12_CeIn3_Spec}(c) and \ref{fig12_CeIn3_Spec}(d). Thus, the respective $T_c^{AFM}$ and $T_c^{PM}$ for AFM and PM are determined as the temperature below which $1/T_1T$ decreases markedly due to the opening of superconducting gap. These results verify that the uniformly coexistent phase of AFM and SC takes place on a microscopic level in the range $P = 2.28 - 2.43$ GPa.
In the PM at $P = 2.43$ and 2.50 GPa, the $1/T_1T$ = const. relation is valid below $T_{FL}\sim 3.2$ K and $\sim 3.5$ K, respectively, which indicates that the Fermi-liquid state is realized. This result is in good agreement with that of the previous resistivity measurement from which the $T^2$ dependence in resistance was confirmed.\cite{Knebel02} Note that the $1/T_1$ for the SC in the PM at $P = 2.50$ GPa follows a $T^3$ dependence below $T_c = 190$ mK, consistent with the line-node gap model characteristic for unconventional HF-SC.

 At $P = 2.17$ GPa, no phase separation occurs below $T_N = 5.5$ K and $1/T_1T = $ const. behavior is observed well below $T_N$. At $P^* = 2.28$ GPa, that is the end point of first-order transition, the $T$ dependence of $1/T_1T$  resembles the behavior at $P = 2.17$ GPa above $T \sim$ 1 K. However, the phase separation into AFM and PM occurs in the small $T$ window between $T_N = 5.2$ K and 3 K. In contrast to the behavior of $1/T_1T$=const. at $P = 2.17$ GPa, the $1/T_1T$ at $P^*= 2.28$ GPa continues to increase in spite of antiferromagnetic spin polarization being induced upon cooling below $T_N$. These results suggest that they are required for the onset of SC to exceed the end point for the first-order transition of AFM and to experience the large enhancement in the low-lying magnetic excitations spectrum below $T_N$. This feature is also seen for the AFM at $P_c = 2.43$ GPa. Some spin-density fluctuations may be responsible for this feature in association with the first-order transition from AFM to PM at $P_c$. In this context, CeIn$_3$ is not in a magnetically soft electron liquid state,\cite{Mathur98} but instead, the relevant magnetic excitations, such as spin-density fluctuations, induced by the first-order transition from AFM to PM might mediate attractive interaction. Whatever its pairing mechanism is at $P = 2.28$ GPa where AFM is realized over the whole sample below $T = 3$ K, the clear decrease in $1/T_1T$ that coincides with the appearance of diamagnetism in $\chi_{ac}$ provide convincing evidence for the uniformly coexistent phase of AFM and SC in CeIn$_3$.

Further evidence for the new type of SC uniformly coexisting with AFM was obtained from the results at $P_c = 2.43$ GPa, as indicated in Fig. \ref{fig13_T1T}(b). Unexpectedly, the magnitude of $1/T_1T=$ const. well below $T_c$ for AFM and PM coincides one another. It should be noted that this behavior of $1/T_1T=$ const. at $P_c=2.43$ GPa  cannot be ascribed to some impurity effect. This is because the $1/T_1$ at $P=2.50$ GPa follows a $T^3$ behavior that is consistent with the line-node gap, nevertheless its $T_c$ goes down from $T_c^{PM}$ = 230 mK at $P_c=2.43$ GPa to 190 mK at $P=2.50$ GPa. This means that the quasi-particle excitations for the uniformly coexistent phase of SC and AFM may be the same in origin as for the phase of SC in PM. How does this happen? It may be possible that both the phases at $P_c$ are  dynamically fluctuating with frequencies smaller than the NQR frequency so as to make each superconducting phase under AFM and PM uniform. In this context, the observed magnetically separated phases and the relevant phases of SC may belong to new phases of matter that may be caused by {\it quantum phase separation through the Josephson coupling}.
\subsection{A new superconducting phenomenon mediated by spin-density fluctuations near the first-order magnetic criticality in CeIn$_3$}
We have provided evidence for the first-order transition at $P_c\sim 2.43$ GPa and the new type of SC uniformly coexisting with the AFM in the range $P^*=2.28$ GPa and $P_c=2.43$ GPa. It has been found that the highest value of $T_c = 230$ mK in CeIn$_3$ is observed for the PM at $P_c$. The present experiments have revealed that this new type of SC uniformly coexisting with the AFM is mediated by a novel pairing interaction in association with the first-order transition. 
In this uniformly coexistent phase of AFM and SC, the $1/T_1T$ measurements have revealed the large enhancement of the low-lying magnetic excitations at the antiferromagnetic state well below $T_N$, being larger than the value for PM, as shown in Figs.\ref{fig13_T1T}(a) and \ref{fig13_T1T}(b). This is because the fluctuations in transverse component of internal fields at the In site are induced by such magnetic density fluctuations even below $T_N$, making low-lying magnetic excitations develop. We propose that {\it the antiferromagnetic spin-density fluctuations, in association with the first-order magnetic criticality, might mediate attractive interaction to form Cooper pairs in CeIn$_3$; this is indeed a new type of pairing mechanism}.
\section{Towards Understanding of Universal Concept for the Superconductivity in Heavy-Fermions Systems}
The SC in HF compounds has not yet been explained from the microscopic point of view, mainly due to the strong correlation effect and the complicated band structures. An essential task seems to identify the residual interaction between quasi-particles through analyzing the effective $f$-band model by choosing dominant bands.\cite{Yanase} Here, we have demonstrated that HF superconductors possess a great variety of ground states at the boundary between SC and AFM with anomalous magnetic and superconducting properties.

A genuine uniformly coexistent phase of AFM and SC has been observed in CeCu$_2$Si$_2$ and CeRhIn$_5$ in the $P$ vs $T$ phase diagram through the extensive and precise NQR measurements under $P$. In other strongly correlated electron systems, the SC appears near the boundary to the AFM. Even though the underlying solid-state chemistries are rather different, the resulting phase diagrams are strikingly similar and robust. This similarity suggests that the overall feature of all these phase diagrams is controlled by a single energy scale. In order to gain an insight into the interplay between AFM and SC, here, we try to focus on a particular theory, which unifies the AFM and SC of the heavy-fermion systems based on an SO(5) theory, because symmetry unifies apparently different physical phenomena into a common framework as all fundamental laws of Nature.\cite{SO(5)review} 
The uniformly coexistent phase diagram of AFM and SC and the exotic SC emerging under strong antiferromagnetic fluctuations could be understood in terms of an SO(5) superspin picture.\cite{kitaoka01,zhang97}

By contrast, CeIn$_3$ has revealed the $P$-induced first-order transition from AFM to PM as functions of pressure and temperature near the boundary at $P_c$.  Unexpectedly, however, the SC is robust even though  the first-order transition takes place at $P_c$ from AFM to PM. The SC for the PM at $P_c$ occurs under the Fermi-liquid state established below $T\sim$ 3.2 K without the development of antiferromagnetic spin fluctuations. The SC uniformly coexisting with the AFM in between $P^*$ and $P_c$ has revealed the mean-field type of transition. The SC revealing the gapless excitations for both AFM and PM at $P_c$ strikingly differ from the SC revealing the line-node gap for the PM in $P > P_c$. The {\it longitudinal} spin-density fluctuations below $T_N$ develop in association with the first-order magnetic criticality from AFM to PM, leading to the onset of the SC uniformly coexisting with the AFM. 
These new phenomena observed in CeIn$_3$ should be understood in terms of a {\it quantum} first-order transition because these new phases of matter are induced by applying pressure. In Fermion systems, if the magnetic critical temperature at the termination point of the first-order transition is suppressed at $P_c$, the diverging magnetic density fluctuations inherent at the critical point from AFM to PM become involved in the quantum Fermi degeneracy region. The Fermi degeneracy by itself generates various instabilities called as the Fermi surface effects, one of which is a superconducting transition. On the basis of a general argument on quantum criticality, it is shown that the coexistence of the Fermi degeneracy and the critical density fluctuations yield a new type of quantum criticality.\cite{Imada} In this context, the results on CeIn$_3$ deserve further theoretical investigations. 

We believe that the results presented here on CeCu$_2$Si$_2$, CeRhIn$_5$ and CeIn$_3$ provide vital clue to unravel the essential interplay between AFM and SC, and to extend the universality of the understanding on the SC in strongly correlated electron systems.
\section{Acknowledgement}
These works have been done in collaboration with K.~Ishida, G.-q. Zheng, G. Geibel, F. Steglich, Y. \={O}nuki and his co-workers. These works were supported by a Grant-in-Aid for Creative Scientific Researchi15GS0213), MEXT and The 21st Century COE Program supported by the Japan Society for  the Promotion of Science. S. K. has been supported by a Research Fellowship of the Japan Society for the Promotion of Science for Young Scientists.

\end{document}